\documentclass[twocolumn,showpacs,preprintnumbers,amsmath,amssymb]{revtex4}

\usepackage{graphicx}

\usepackage{dcolumn}

\usepackage{bm}

\begin{document}

\title{Evidence of two-dimensional macroscopic quantum tunneling of a current-biased DC-SQUID}

\author{F. Balestro$^{1}$, J. Claudon$^{1}$, J. P. Pekola$^{1,2}$, and O. Buisson$^{1}$}

\affiliation{$^{1}$Centre de Recherches sur les Tr\`es Basses
Temp\'eratures, laboratoire associ\'e \`a l'Universit\'e Joseph
Fourier, C.N.R.S., BP 166, 38042 Grenoble-cedex 9, France\\
$^{2}$Low Temperature Laboratory, Helsinki University of
Technology, P.O. Box 2200, 02015 HUT, Finland}

\begin{abstract}
    The escape probability out of the superconducting state of a hysteretic DC-SQUID
    has been measured at different values of the applied magnetic flux.
    At low temperature, the escape current and the width of the probability distribution are temperature independent but
they depend on flux. Experimental results do not fit the usual one-dimensional (1D) Macroscopic Quantum Tunneling (MQT)
   law but are perfectly accounted for by the two-dimensional (2D) MQT behaviour as we propose here. Near zero flux,
    our data confirms the recent MQT observation in a DC-SQUID \cite{Li02}.

\end{abstract}


\maketitle

Quantum dynamics of a current-biased Josephson junction (JJ) have been extensively studied during the past two decades
\cite{Leggett92}. The phase difference across the JJ obeys the same dynamics as the
position of a fictitious particle
moving in a one dimensional tilted washboard potential, whose average slope is
proportional to the current bias. For bias below the critical current, the particle is trapped in a local minimum. It can escape
out by Thermal Activation (TA) or by
Macroscopic Quantum Tunneling (MQT). Many experiments have demonstrated these two escape regimes.
Quantized energy levels and resonant activation have been observed by applying microwaves \cite{Leggett92,Esteve86}.
Very recently Rabi oscillations and coherent temporal oscillations were observed in a current biased JJ
\cite{Yu02,Martinis02} or in a Cooper pair transistor coupled to a biased JJ\cite{Vion02}. Experiments on RF-SQUID have also clearly demonstrated quantum
dynamics in form of MQT \cite{Leggett92}, resonant tunneling or quantum superposition of states \cite{Friedman00}.

By contrast and surprisingly, very few studies have been performed in a current biased DC-SQUID in the quantum regime.
The dynamics of a DC-SQUID are described by two degrees of freedom and new phenomena can arise.
The fictitious particle moves in this case in a two-dimensional (2D) potential. At zero magnetic field it has been predicted
that escape out of a local minimum can occur
via two different trajectories \cite{Ivlev87,Morais94}. At non zero magnetic field, MQT
was derived only in the limit of small inductance of the loop \cite{Chen86}, where the SQUID is nearly equivalent to a
single junction with a flux-dependent critical current. In this case, the dynamics are one dimensional.
To our knowledge only three previous experiments were reported in the past discussing escape from a current-biased DC-SQUID.
In the TA regime good agreement with 2D transition state theory \cite{Brinkman57} has been obtained \cite{Seguin92}.
Similar results were obtained in an RF-SQUID \cite{Han89}.
In the quantum regime no experimental evidence of two-dimensional behaviour has been demonstrated
up to now.
Sharifi et al. \cite{Sharifi88} claimed MQT observations but neither the amplitude of quantum
fluctuations nor the TA regime were understood. Recent measurements \cite{Li02} have clearly observed MQT and TA regimes
in a DC-SQUID {\sl at zero magnetic field}. 2D behaviour could, however, not be observed because at zero flux
the escape rate of their SQUID behaves exactly as that of a single Josephson junction, ie. a 1D system.
Understanding the dynamics of a hysteretic DC-SQUID is very important in interpreting some
recent quantum experiments \cite{VanderWal00,Saito02}, or a recently
proposed experiment of a one shot quantum measurement in a superconducting charge qubit using a DC-SQUID \cite{Buisson03}.

In this letter, we report on an escape measurement in a DC-SQUID both in the thermal and in the quantum regime.
In contrast to till now reported results on MQT, 1D tunneling escape formula can not describe our experimental data.
Since no theory has described MQT in a DC-SQUID in the 2D regime at non zero magnetic flux,
we propose a method based on the zero magnetic field result derived in Ref. \cite{Ivlev87} but taking into account the
flux dependence of the 2D potential. Our experimental data support quantitatively this approach.

The current-biased DC-SQUID consists of two Josephson junctions in parallel in an inductive superconducting loop.
The whole system has two degrees of freedom described by $\varphi_{1}$ and $\varphi_{2}$, the phase difference
across the two junctions. Assuming identical JJs, each having a critical current equal to $I_{0}$ and a capacitance equal to
$C_{0}$, the dynamics of the system can be treated
as that of a fictitious particle moving in a two dimensional potential given by \cite{Tesche77}
   \begin{equation}
       U(x,y)=U_{0}\left(-sx-\cos{x}\cos{y}-\eta{s}y+b(y-\pi\Phi/\Phi_{0})^{2}\right)
    \label{2Dpotential}
    \end{equation}
where $U_{0}=\Phi_{0}I_{0}/\pi$ is the sum of the Josephson energy
of the two JJs, $s=I_{DC}/(2I_{0})$,
$x=(\varphi_{1}+\varphi_{2})/2$, $y=(\varphi_{1}-\varphi_{2})/2$,
$b=\Phi_{0}/(2\pi{L}I_{0})$ and $I_{DC}$ is the bias current. The
asymmetry of inductances is parametrized by $\eta=(L_{2}-L_{1})/L$
where $L_{1}$ and $L_{2}$ are the inductances of the two branches
and $L=L_{1}+L_{2}$ is the inductance of the SQUID loop. $\Phi$ is
the applied flux and $\Phi_{0}=h/2e$ is the flux quantum.

The potential presents valleys and mountains with local minima separated by saddle points along which the particle can
escape.
The barrier height between the minima and the saddle points depends on the bias current and magnetic flux and vanishes at
the critical current $I_{c}$.
$I_{c}$ follows the usual flux dependence with $\Phi_{0}$ periodicity.
In the following we consider $b\sim 1$  close to that of our
   DC-SQUID. In this limit, there exists only
    one optimal trajectory that passes through the saddle point. Therefore we will not discuss escape out of a local minimum
through two different trajectories \cite{Ivlev87,Morais94}. Yet the 2D signature still
exists because the optimal trajectory is not a straight line along the $x$-direction but follows a path in the landscape
with the tilt in the potential.

\begin{figure}
\resizebox{.4\textwidth}{!}{\includegraphics{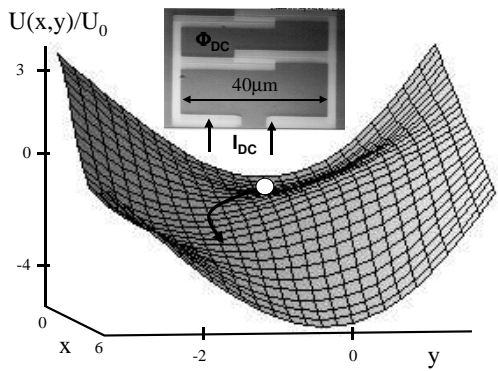}}
\caption{\label{Potential}Bidimensional potential of our SQUID
which has parameters $s=0.7$, $\Phi/\Phi_{0}=-0.27$ and $b=0.6$
and $\eta=-0.26$. The particle and its escape path are drawn as
illustration. Inset: photograph of the type of a DC-SQUID used in
the experiment.}
\end{figure}

In the limit $s\lesssim{s_{c}}$ where $s_{c}=I_{c}/(2I_{0})$ is the normalized effective critical current
    of the DC-SQUID, we can expand the potential (Eq. (\ref{2Dpotential})) to the third order. The escape occurs in a
    current and flux dependent angle $\theta$
    with respect to the $x$-axis. Along this direction, we obtain the
potential
  barrier height, $\Delta{U}=\Delta{U_{0}}\sqrt{{s_{c}\cos^{3}{\theta}}/{u}}(1+\eta\tan\theta)^{3/2}$, and
  the plasma frequency, $\omega_{p}=\omega_{p0}({u\cos{\theta}}/{s_{c}})^{1/4}(1+\eta\tan\theta)^{1/4} $
  where $u$ is the third derivative of the potential along the escape direction \cite{Seguin92}.
  The quantities $\omega_{p0}=2^{1/4}(\pi{I_{c}}/\Phi_{0}C_{0})^{1/2}(1-s/s_{c})^{1/4}$ and
    $\Delta{U_{0}}=(4\sqrt{2}/3)(\Phi_{0}/2\pi)I_{c}(1-s/s_{c})^{3/2}$ are the usual  plasma frequency and potential height, respectively,
    of a single JJ with an effective critical current $I_{c}$ and a capacitance $2C_{0}$.
   The 2D character of the escape is exhibited in these expressions by the renormalization of the plasma frequency and the
potential barrier height of a single JJ. Therefore the ratios
$\omega_{p}/\omega_{p0}$ and $\Delta{U}/\Delta{U_{0}}$ show deviations from 1D behavior.

Using the previous results of the renormalized plasma frequency and potential height,
the escape rate out of a 2D potential well, in the TA regime, $\Gamma^{TA}$, is predicted to be \cite{Brinkman57}
    \begin{equation}
    \Gamma^{TA}=\frac{\omega_{p}}{2\pi}\frac{\omega_{w\bot}}{\omega_{s\bot}}\exp\left(-\frac{\Delta{U}}{k_{B}T}\right)
   \label{escaperateTA}
   \end{equation}
where ${\omega_{w\bot}}$ and ${\omega_{s\bot}}$ are respectively the transverse frequency modes in the well and at
    the saddle point \cite{Seguin92}. We numerically analyzed the modulation of the escape current as a function of flux
    in the TA regime taking into account the transverse frequency modes in the escape rate. For our SQUID parameters, the
    deviation of ${\omega_{w\bot}}/{\omega_{s\bot}}$ from unity has been
    estimated to be $2\%$ at zero flux and it quickly decreases to zero when increasing $\mid\Phi/\Phi_{0}\mid$.
    Since this small error appears only in the prefactor of Eq. (\ref{escaperateTA}),
we will assume in the following that ${\omega_{w\bot}}/{\omega_{s\bot}}=1$.

At low temperature, the escape is dominated by quantum tunneling. This regime has been
    theoretically studied at zero flux in \cite{Ivlev87, Morais94} discussing the splitting of the two escape trajectories,
or flux dependence in the low inductance limit ($b>>1$) \cite{Chen86}. There is so far, to our knowledge, no theoretical
    treatment of flux dependence of 2D MQT, ie. in the regime $b\lesssim{1}$. Therefore we propose to use the MQT escape rate at zero
magnetic flux \cite{Ivlev87, Li02} in the limit $b\approx 1$.
  We introduce in this escape rate, $\Gamma^{MQT}$, the 2D behaviour by taking
    the plasma frequency $\omega_{p}$ and the potential barrier height $\Delta{U}$ along the escape direction, analogously
to what was done in the TA regime earlier \cite{Seguin92}:
    \begin{equation}
    \Gamma^{MQT}=f_{2D}\frac{\omega_{p}}{2\pi}\sqrt{864\pi{\Delta{U}/\hbar\omega_{p}}}
    \exp{(-36\Delta{U}/5\hbar\omega_{p})}.
    \label{escaperateMQT}
\end{equation}
     This 2D MQT rate thus depends on the magnetic flux and on the bias-current.
In the following, because  $b\approx 1$, two escape trajectories are impossible and $f_{2D}\simeq 1$. Note, however, that
the 2D behaviour remains through
$\Delta{U}$ and $\omega_{p}$. At zero magnetic flux  $\Gamma^{MQT}$ equals that of a single JJ of critical current $2I_{0}$
and capacitance $2C_{0}$.

\begin{figure}
\resizebox{.35\textwidth}{!}{\includegraphics{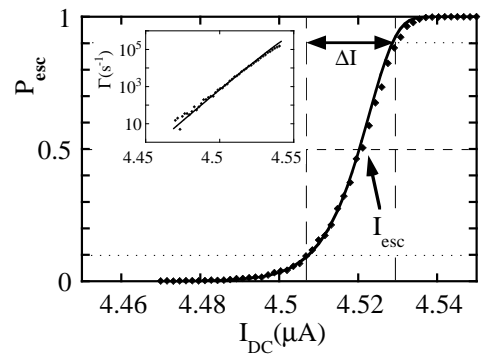}}
\caption{\label{EscapePro} Measured escape probability and escape
rate (inset) as functions of the bias current at
$\Phi/\Phi_{0}=0.07$, $T=42$ mK and $\Delta{t}=50$ $\mu$s. The
dots are experimental data and the lines the MQT prediction of Eq.
(\ref{escaperateMQT}). Values of $I_{esc}$ and $\Delta I$ are
indicated by dashed lines. }
\end{figure}

The escape probability $P_{esc}$ from the zero voltage state during time $\Delta{t}$
    is a function of the amplitude of the bias current, the flux and temperature.
Measured escape probability  using current DC-pulses of duration $\Delta{t}=50$ $\mu$s is plotted in Fig. \ref{escaperateMQT}.
Each point corresponds to 4000 current pulses.
 By increasing the bias current, $P_{esc}$ starts from zero far below the critical current where the barrier
 is high, and it becomes unity at currents close to and above the critical one.
In the following the escape current $I_{esc}$ is defined as the bias current corresponding to $P_{esc}=0.5$
    and the width of the escape probability distribution $\Delta I$ as the difference between the currents at $P_{esc}=0.9$
and $P_{esc}=0.1$.
    The measured escape rate is extracted from $\Gamma(I_{DC})=-\ln(1-P_{esc}(I_{DC}))/\Delta{t}$
   and it ranges from about 10 s$^{-1}$ to 10$^{5}$ s$^{-1}$ (inset of Fig. \ref{EscapePro}).

\begin{figure}
\resizebox{.35\textwidth}{!}{\includegraphics{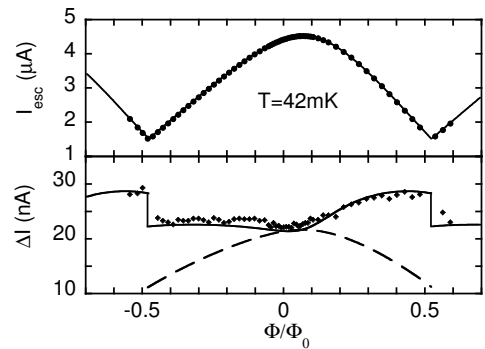}}
\caption{\label{escapevsflux} Measured escape current and the
width of the probability distribution against external applied
flux at low temperature (dots). 1D MQT and 2D MQT prediction (Eq.
\ref{escaperateMQT}) are shown respectively by dashed and solid
lines.}
\end{figure}

The sample is a DC-SQUID consisting of two $Al/AlO_x/Al$ tunnel junctions
    obtained by e-beam lithography and shadow evaporation (inset of Fig. \ref{Potential}).
    The area of each JJ is about 11 $\mu$m$^{2}$. The tunnel resistance of the SQUID, $R_{T}\simeq 80$ $\Omega$, and
    the superconducting gap $\Delta=230$ $\mu$eV
    are deduced from the dissipative branches of the IV characteristic
and yield $I_{0}\simeq 2.3$ $\mu$A
    using the Ambegaokar-Baratoff
    relation assuming two identical JJs.
    From MQT of a single JJ with the same area, we deduced $C_{0}=0.46$ pF, which agrees with the typical specific
    capacitance of about 50 fF/($\mu$m)$^2$ for an $AlOx$ tunnel junction. The
rectangular SQUID loop is about 10 $\mu$m by
40 $\mu$m giving
    an estimated magnetic inductance of about 70 pH. The kinetic inductance of the SQUID,  deduced
    from the thin film resistivity, $\rho\simeq 5.3\mu\Omega$cm,
    is coarsely estimated to be about 120 pH. The total inductance $L$, the sum of the kinetic and magnetic inductances,
    is therefore about 190 pH.
    An inductance asymmetry exists because the two SQUID branches have different lengths. We estimate $\eta\sim -0.2$
    by assuming proportionality between length of a branch and inductance.
    The immediate electromagnetic environment of the DC-SQUID is determined by two 20 nm thin, 1 $\mu$m wide and 700 $\mu$m
long superconducting aluminium wires terminated by two large pads for wire bonding.
    The estimated total inductance of the in-situ wires, $L_{e}=4.6$ nH,  results again of the sum of 1.2 nH magnetic inductance of
    the coplanar strips and of 3.4 nH kinetic inductance of the wires.
        Although these SQUID parameters have been independently estimated, the accuracy in determining them this way is not
sufficient.
    Therefore we extracted the inductance of the DC-SQUID $L=244$ pH
    and its asymmetry $\eta=-0.26$, the zero-flux,
    the critical current $I_{0}=2.33$ $\mu$A and the capacitance $C_{0}=0.46$ pF from the flux modulation characteristics of the escape current in the low temperature
regime (see Fig. \ref{escapevsflux}). These parameters are in very good agreement with the rough
    estimates given above.
   Our $b=0.6$ is much smaller than
    $b\approx{3}$ of the earlier works \cite{Sharifi88,Li02} on MQT in a DC-SQUID, indicating that
    the fictitious particle is much more affected by the two-dimentionality of the potential.

The sample is enclosed in a cavity with resonances above 15 GHz, and it is anchored to the mixing chamber of the
dilution
    refrigerator. This cavity is
    enclosed in a low temperature filter \cite{Balestro03} consisting of thermocoax microwave
    filters \cite{Zorin95} and $\Pi$-filters. The attenuation of the DC-lines is estimated to be at least
    200 dB above 1 GHz at the mixing chamber temperature.
    Thermocoax filters were also introduced from 1.5 K to the low temperature filter, and low pass LC filters
    were used at room temperature.
    Current bias and voltage probe lines of the sample are separated from the computer using differential amplifiers. A
superconducting shield was inserted inside the vacuum jacket of the cryostat and $\mu$-metal
    surrounds the dewar to protect the DC-SQUID from external flux noise.
    Using this measurement set-up, we separately measured a width of $\Delta I=11$ nA on a 1 $\mu$A Josephson junction
    in its MQT regime.

Figure \ref{EscapePro} shows escape probability and escape rate at $\Phi/\Phi_{0}=0.07$ and $T=42$ mK.
At this particular flux value, the escape current is maximum (see Fig. \ref{escapevsflux}) and
the ratios $\omega_{p}/\omega_{p0}$ and $\Delta{U}/\Delta{U_{0}}$ are equal to unity. Therefore the SQUID
behaves as a single JJ. We can perfectly fit our data using
  1D-MQT predictions (continuous line in Fig. \ref{EscapePro}) without free parameters which demonstrates that
  1D-MQT theory describes the behaviour of a DC-SQUID at the maximum escape current. This
  first result confirms the recent MQT observation at zero flux performed by Li $et$ $al.$ \cite{Li02}.
  It does not, however, demonstrate 2D escape behaviour.

The escape current and the width versus applied external flux are
directly extracted from the measured switching current
distribution and they are plotted in Fig. \ref{escapevsflux} in
the MQT regime at T= 42mK . The width is periodic in flux with
periodicity $\Phi_{0}$. It is minimum near $\Phi/\Phi_{0}=0.07$
and increases to reach maximum at $\Phi/\Phi_{0}=\pm 0.5$. There
is asymmetry between negative and positive flux values and
discountinuities at $\Phi/\Phi_{0}=\pm 0.5$. The usual 1D
tunneling model cannot account for the main feature of our
results. Namely the 1D model predicts a decrease of the width,
$\Delta I \sim (2I_{0}s_{c})^{3/5}$ in the MQT regime (dashed
line in Fig. \ref{escapevsflux}) which is contrary to our
observation. Our proposed 2D formula in the MQT regime suggests
increase of the width and it agrees with our data perfectly also
quantitatively. It has no free parameters since all of them were
already fixed by the escape current characteristics. We claim that
these results are the first experimental observation of 2D escape
from the zero voltage state of a DC-SQUID in its MQT regime.

\begin{figure}
\resizebox{.35\textwidth}{!}{\includegraphics{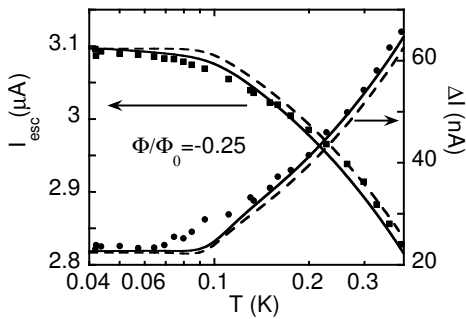}}
\caption{\label{WidthT} Measured escape current and width of
escape probability distribution $\delta I$ vs. temperature
(symbols) at $\Phi/\Phi_{0}=-0.25$ together with
predictions when including (continuous line) or excluding (dashed line) the influence of the environment. We assume that the total escape rate is the sum of $\Gamma_{MQT}$ and 
$\Gamma_{TA}$.}
\end{figure}

The escape current and width against temperature are shown in Fig.
\ref{WidthT} for $\Phi/\Phi_{0}=-0.25$. The escape current
increases as temperature drops and saturates at low temperature.
 The width decreases with temperature and saturates at
low temperature.
These features are perfectly accounted for by Eq. (\ref{escaperateMQT}) in the MQT regime and by
Eq. (\ref{escaperateTA}) in the TA regime (dashed lines). Yet the model
seems to underestimate
the measured width in the thermal regime by about 10 \%, indicating that influence of the environment must be taken into account to better fit with the data.
Finally, the observed cross-over temperature ($\simeq 80$ mK) between MQT and TA is consistent with that predicted by the
well-known formula $\hbar\omega_{p}/(2\pi k_{B})$ \cite{Leggett92}.

In order to estimate the effect of the environment on the escape process,
we take into account the quadratic current fluctuations $\langle\delta I^{2}\rangle$. Since the SQUID is connected to the external
circuit through inductive wires on the chip,
we assume the electrical environment to be the inductance $L_{e}$ in series with an external resistor $R_{ext}$. For temperatures
$k_{B}T>\hbar R_{ext}/L_{e}$, the current fluctuations are given by $\langle\delta I^{2}\rangle=k_{B}T/L_{e}$.
For $k_{B}T<\hbar \omega _{p}$
thermal fluctuations are slow as compared to the SQUID dynamics, and escape rate can be corrected by
 $\Gamma^{TA}_{env}=\Gamma^{TA}\exp(({\partial\Delta{U}}/{\partial{I_{DC}}}){\langle\delta I^{2}\rangle}/{2(k_{B}T)^{2}})$
    in the TA regime
and
$\Gamma^{MQT}_{env}=\Gamma^{MQT}\exp(({\partial{(\Delta{U}/\hbar\omega_{p})}}/{\partial{I_{DC}}}){\langle\delta I^{2}\rangle}/{2})$
in the MQT regime following the idea in \cite{Martinis88}.
Using  the estimated wires inductance, our data perfectly are fitted both in the quantum and in the TA regime
by the fluctuation corrected expressions above (continuous line in Fig. \ref{WidthT}). We notice that like in the recent work in \cite{Li02}
the measured $\Delta I$ crosses over from MQT to TA much more gradually than what we predict.

In summary, we have shown for the first time evidence of two dimensional macroscopic quantum escape
of a hysteretic DC-SQUID from its zero voltage state. Moreover, we have proposed a way to theoretically
analyze the flux dependence of escape in the MQT
regime, and we show that our data perfectly fit this model, also in the TA regime with the same values of the circuit parameters.

We thank K. Hasselbach, F. Hekking, Ph. Lafarge, L. L\'evy and A. Niskanen for
useful discussions. This work was supported by the French ACI
program. JP acknowledges support from CNRS and Joseph
Fourier University.


\begin{thebibliography}{99}

\bibitem{Leggett92}
{\sl Quantum Tunneling in Condensed Media}, Modern Problems in Condensed Matter Sciences, Vol. 34, edited by Yu. Kagan and
A. J. Leggett
(Elsevier Science Publishers, 1992).
\bibitem{Yu02}
Y. Yu, {\em et al.}, Science {\bf 296}, 889 (2002).
\bibitem{Martinis02}
John M. Martinis, S. Nam, J. Aumentado, and C. Urbina, Phys.  Rev.  Lett. {\bf 89}, 117901 (2002).
\bibitem{Esteve86}
D. Esteve, M. H. Devoret, and J. M. Martinis,Phys. Rev. B {\bf 34}, 158 (1986).
\bibitem{Vion02}
D. Vion, {\em et al.}, Science {\bf 296}, 886 (2002).
\bibitem{Friedman00}
J. R. Friedman, {\em et al.}, Nature {\bf 406}, 43 (2000).
\bibitem{Ivlev87}
B. I. Ivlev and Yu.  N. Ovchinnikov, Sov. Phys. JETP {\bf 66}, 378 (1987).
\bibitem{Morais94}
C. Morais Smith, B. Ivlev, and G. Blatter, Phys. Rev. B {\bf 49}, 4033 (1994).
\bibitem{Chen86}
Y.-C. Chen, J. Low Temp. Phys. {\bf 65}, 133 (1986).
\bibitem{Brinkman57}
H. C. Brinkman, Physica {\bf 22}, 149 (1957).
\bibitem{Seguin92}
V. Lefevre-Seguin, {\em et al.}, Phys. Rev. B {\bf 46}, 5507 (1992).
\bibitem{Han89}
S. Han, J. Lapointe, and J. E. Lukens, Phys.  Rev.  Lett. {\bf 63}, 1712 (1989).
\bibitem{Sharifi88}
F. Sharifi, J. L. Gavilano, and J. van Harlingen, Phys.  Rev.  Lett. {\bf 61}, 742 (1988).
\bibitem{Li02}
S. Li, {\em et al.}, Phys.  Rev.  Lett. {\bf 89}, 98301 (2002).
\bibitem{VanderWal00}
C. H. van der Wal, {\em et al.}, Science {\bf 290}, 773 (2000).
\bibitem{Saito02}
S. Saito, {\em et al.},
    in {\sl Macroscopic quantum coherence and computing}, p. 137, edited by
    J. Pekola, B. Ruggiero and P. Silvestrini (Kluwer Academics, New York, 2002).
\bibitem{Buisson03}
O. Buisson, F. Balestro, J. P. Pekola, and F. W. J. Hekking, Phys. Rev. Lett. {\bf 90}, 238304 (2003).
\bibitem{Tesche77}
C. D. Tesche and J. Clarke, J. Low Temp.  Phys. {\bf 29}, 301 (1977).
\bibitem{Zorin95}
A. B. Zorin, Rev. Sci. Instrum. {\bf 66}, 4296 (1995).
\bibitem{Balestro03}
F. Balestro, PhD Thesis, Universite Joseph Fourier, Grenoble, France (2003).
\bibitem{Martinis88}
J. M. Martinis and H. Grabert, Phys. Rev. B {\bf 38}, 2371 (1988).
\end{thebibliography}
\end{document}